\documentclass[onecolumn]{IEEEtran}
\IEEEoverridecommandlockouts
\usepackage{cite}
\usepackage{amsmath,amssymb,amsfonts,amsthm}
\usepackage{graphicx}
\usepackage{textcomp}
\usepackage{xcolor}
\usepackage[utf8]{inputenc}
\usepackage[english]{babel}
\usepackage{siunitx}
\usepackage{booktabs}
\usepackage{makecell}
\usepackage{ltxtable,xcolor}
\usepackage{bm}
\usepackage{svg}
\usepackage{caption}
\usepackage{subfig}
\usepackage{algorithm,algorithmicx,algpseudocode}

\usepackage [english]{babel}
\usepackage [autostyle, english = american]{csquotes}
\MakeOuterQuote{"}

\graphicspath{ {./images/} }

\newcommand{\etal}{{et al}.\@ }
\definecolor{mygray}{gray}{0.30}
\setcellgapes{3pt}

\def\BibTeX{{\rm B\kern-.05em{\sc i\kern-.025em b}\kern-.08em
    T\kern-.1667em\lower.7ex\hbox{E}\kern-.125emX}}

\theoremstyle{definition}
\newtheorem{remark}{Remark}

\begin{document}

\title{Distributed Complementary Fusion for Connected Vehicles}

\author{\IEEEauthorblockN{James~Klupacs, Amirali~Khodadadian~Gostar, Alireza~Bab-Hadiashar, Jennifer Palmer, and~Reza~Hosseinezhad}\\
\IEEEauthorblockA{\textit{RMIT University}\\
Melbourne, Australia}}

\maketitle

\begin{abstract}
We present a random finite set-based method for achieving comprehensive situation awareness by each vehicle in a distributed vehicle network. Our solution is designed for labeled multi-Bernoulli filters running in each vehicle. It involves complementary fusion of sensor information locally running through consensus iterations. We introduce a novel label merging algorithm to eliminate double counting. We also extend the label space to incorporate sensor identities. This helps to overcome label inconsistencies. We show that the proposed algorithm is able to outperform the standard LMB filter using a distributed complementary approach with limited fields of view.
\end{abstract}

\begin{IEEEkeywords}
 Random Finite Sets, Intelligent Transport Systems, Multi-Object Tracking, Information Fusion, Distributed Algorithms
\end{IEEEkeywords}

\section{Introduction}
The emergence of \textit{connected vehicles} in near future, as a dominant part of intelligent transport systems, is inevitable. At any time, there may be tens of vehicles in an area, with an unknown number of potential obstacles such as roadside debris or pedestrians. Whether the connected vehicle is driven, semi-autonomous, or fully-autonomous, achieving \textit{comprehensive} situational awareness is essential for provision of effective driver-assist or self-driving commands, ensuring safety and efficiency.

In practice, the connected vehicles are equipped with sensors that each have limited field-of-view (FoV). Hence, to achieve \textit{comprehensive} situational awareness, a \textit{complementary} sensor fusion algorithm should be in place.  

Our in proposed distributed vehicle network structure, we assume that situational awareness is acquired from a statistical model; a multi-object density to be more precise. Having the density, we can then estimate the number of objects and their states (e.g. location, speed). In addition, we can measure the \textit{accuracy} of or \textit{confidence} in the reliability of the estimates, by investigating statistical properties of the density. 

Assuming that a multi-object posterior is formed in each vehicle node of the network as the result of running a multi object tracking (MOT) filter. The recently developed MOT filters under the random finite set (RFS) framework~\cite{mahler2014advances,9074662,GOSTAR2020107278} have shown promising results in various applications such as track-before-detect visual tracking~\cite{s20030929}, sensor management in target tracking applications~\cite{PANICKER2020107451,8455829} and information fusion~\cite{9311857}. We choose the labeled multi-Bernoulli (LMB) filter as the filter of choice. An LMB posterior is fully characterized by a number of possibly existing objects each associated with label, a probability of existence and a single-object density for the object state.

Distributed networks are the preferred method of computing these problems as they are infinitely scalable with nodes, that is they scale in a linear fashion. This is in contrast to centralized networks which although are simpler to implement, and allow for the optimal solution to be computed the cost scales exponentially. In an ITS scenario the only possible method to potentially achieve real time networks is to use distributed computing. Much work has been done in the area of distributed RFS filters~\cite{li2018distributed,9136678,Battistelli2013508,9478329} however they have generally been limited to using Kullback-Leibler Averaging (KLA) or consider unlimited FoVs and have not been applied to the use case of connected vehicles with their challenges of complex motion, large sensor networks and limited FoVs.


The distributed network of connected vehicles has a \textit{dynamic} nature, in the sense that its structure changes with time.
 Note that in this dynamic network, each node itself is among the objects that need to be detected and its state to be estimated by others, i.e. it is part of situational awareness.  The dynamic changes mean that there is no way of guaranteeing object labels computed and propagated within the MOT filter in one node will be the same at another~\cite{8060598}. For example, a vehicle that is already labeled in the one node, is given a different (newly born label) when entering the FoV of another node. In addition, complementary fusion does not intrinsically avoid double-counting of information. 

The main contributions of this paper are: (1)  a distributed implementation of the complementary fusion method, which was previously proposed in~\cite{GostarAmiraliKhodadadian2018CMIF}, revised for use in a time-varying connected vehicle network, (2) a merging solution to remove double-counting that is inevitable with complementary fusion, and (3) a novel solution to overcome the label inconsistency present in distributed network. 

\section{Background}

\subsection{Network model}
In this paper we consider a sensor network denoted by $(\mathcal{N},\mathcal{A})$ where $\mathcal{N}$ is the set of nodes and $\mathcal{A}\subseteq\mathcal{N}\times\mathcal{N}$ represent the connections between each node. Two nodes $i$ and $i'$ are connected, that is able to communicate and exchange information, if $(i,i')\in\mathcal{A}$. Each node $i\in\mathcal{N}$ has a set of neighbors (including itself) denoted by $\mathcal{N}^i\triangleq\{i'\in\mathcal{N}:(i,i')\in\mathcal{A}\}$. The above sensor network is fully distributed where each node $i$ can only communicate directly with its neighbors. All operations (computation, measurement acquisition) occur locally at each node, without knowledge of the entire network.

\subsection{Information model}  
At each node, the local sensor measurements are processed via a statistical filter, resulting a multi-object density denoted by $\bm{\pi}_{i,k}$ where $k$ is an index to discrete time of measurement acquisition. Situational awareness is achieved by estimating the number and states of various objects (with labels) in the scene from this multi-object density. Being a statistical model enables us to also assess the \textit{confidence} in the resulting situational awareness (the estimates) too. 

We note that different nodes (vehicles) in the network may be equipped with different types of sensors with various modalities, resolutions and sensitivities. The locally computed multi-object density plays the role of \textit{common information representation} that is independent form sensor types and modalities. Importantly, there are application-specific (non-measurement) information such as statistical models for object state transitions (motion model), object birth and object death that are incorporated into the resulting multi-object density via applying the statistical filter.

In the network, each node communicates it locally formed multi-object density to its neighbors, and in-turn, receives theirs. The received densities are then fused to acquire a more informed density that is anticipated to return more reliable situational awareness.
 
\subsection{LMB density}
The multi-object density of choice in this work is a particular RFS density called \textit{labeled multi-Bernoulli} (LMB). This formulation approximates an RFS with the ensemble of multiple \textit{possibly and independently existing} objects. Each object has its label, and is parametrized by its probability of existence and the multivariate density of its state vector. The LMB density is entirely characterized by the parameters of its (possibly existing) objects.

In the context of distributed fusion in connected vehicle networks, at each vehicle node $i$, the LMB density that is locally acquired at time $k$, is parametrized as:
\begin{equation}
	\bm{\pi}_{i,k} \sim \left\{\left(r_{k,i}^{(\ell)}\,,\,p_{k,i}^{(\ell)}(\cdot)\right)\right\}_{\ell\in\mathbb{L}_{k}}
	\label{eq:LMB_parametrization}
\end{equation}
where $\ell$ denotes a single-object label. In addition, $r^{(\ell)}$ and $p^{(\ell)}(\cdot)$ denote the probability of existence and the multivariate density of the single-object state vector, for the object labeled $\ell$, respectively. $\mathbb{L}_k$ denotes the space of all object labels discovered and tracked up to time $k$.

This paper considers the sequential Monte Carlo (SMC) approximation of the single-object densities. Each multivariate density, $p_{k,i}^{(\ell)}(\cdot)$ is represented by particles and their weights as follows:
\begin{equation}
	p_{k,i}^{(\ell)}(\cdot) \approxeq 
	\sum_{j=1}^{J_{k,i}^{(\ell)}} w_{k,i,j}^{(\ell)}\,\delta(x-x_{k,i,j}^{(\ell)}) 
	\label{eq:particles-representation}
\end{equation}
where $J_{k,i}^{(\ell)}$ denotes the number of particles which are themselves denoted each by $x_{k,i,j}^{(\ell)}$ and their weights, each by $w_{k,i,j}^{(\ell)}$.

\begin{remark}
	Given the multi-object density as an LMB given by~\eqref{eq:LMB_parametrization} and~\eqref{eq:particles-representation}, situational awareness can be simply achieved by: (1) considering each object whose probability of existence exceeds a large threshold (e.g. 0.90) as existing, and (2) estimating its state as the weighted average of its particles (this is indeed an Expected A Posteriori - EAP - estimate).
\end{remark}

\subsection{LMB filter}

The parameters of the LMB posterior are propagated through the prediction and update steps of an LMB filter~\cite{Reuter20143246}. At each time $k$, the current LMB density $\bm{\pi}_{i,k}$ parameterized by~\eqref{eq:LMB_parametrization} is input as \textit{prior} to the prediction step. 

In prediction step, the state of each possibly existing object is propagates according to a single-object state transition model, also called \textit{motion model}. The possible exit of each object from the scene (FoV of the vehicle/sensor node in the network) is also taken into consideration via a particular parameter, called \textit{probability of survival}, which is commonly defined as a function of the object state. More importantly, the set of possibly existing objects with propagated states is appended with the set of \textit{possibly born} objects at time $k$. This set formulates the \textit{birth process} and normally includes a number of components with small probabilities of existence and densities that are peaked around possible points of entry to the scene. Each element of this set is labeled $(k,m)$ where $k$ denotes the time of birth and $m$ is an index. For instance, if there are four birth objects, their labels are $\mathbb{B}_k = \left\{(k,1), \ldots, (k,4)\right\}$. Thus, the space of labels is extended as well to $\mathbb{L}_{k+1} = \mathbb{L}_k \cup \mathbb{B}_k$.

The update step inputs the outcome of the prediction step (predicted LMB). Then, Bayes' rule is applied to incorporate the information \textit{measured by sensors}. Uncertainties in terms of false negatives (missing an object and returning no measurements for it) and false positives (false measurements that are not associated with any object - also clutter measurements) are modeled by a state-dependent probability of detection and an RFS model assumed for clutter measurements (normally a Poisson RFS), respectively.

Details of prediction and update steps and their derivations can be found in~\cite{Reuter20143246}. The resulting LMB density is the new posterior, denoted by:
\begin{equation}
	\bm{\pi}_{i,k+1} \sim \left\{\left(r_{k+1,i}^{(\ell)}\,,\,p_{k+1,i}^{(\ell)}(\cdot)\right)\right\}_{\ell\in\mathbb{L}_{k+1}}.
	\label{eq:LMB_posterior}
\end{equation}

\begin{remark}
	\label{rem:pruning_and_resampling}
	For the sake of computational tractability, the components with very low probability of existence are \textit{pruned}. Additionally, the particles of each remaining object are \textit{resampled} to avoid particle deprivation.
\end{remark}

\section{Distributed Information Fusion}

In order to achieve comprehensive situational awareness in each vehicle node, we suggest a distributed information fusion scheme that is comprised of two major components: complementary fusion and consensus iterations.

\subsection{Complementary fusion}
In each vehicle node $i$ at time $k$, all the locally acquired information are  encapsulated in the posterior $\bm{\pi}_{i,k}$. Information fusion occurs by fusing this locally generated posterior with all those communicated from neighboring nodes. We use the complementary fusion scheme first devised by Gostar~\etal in~\cite{GostarAmiraliKhodadadian2018CMIF}. In this scheme, the fused posterior is density of the union of all the posteriors:
\begin{equation}
	\bar{\bm{\pi}}_{i,k} \sim \bigcup_{i'\in\mathcal{N}^i \cup \{i\}}  \left\{\left(r_{k,i'}^{(\ell)}\,,\,p_{k,i'}^{(\ell)}(\cdot)\right)\right\}_{\ell\in\mathbb{L}_{k}}
\end{equation}
approximated by an LMB. The resulting LMB parameters are then given by:

\begin{eqnarray}
	\bar{r}_{k,i}^{(\ell)} & = & 
	\sum_{i'\in\mathcal{N}^i \cup \{i\}} \varrho_{k,i'}^{(\ell)}\,\bigg/\,\bigg[1+ \sum_{i'\in\mathcal{N}^i \cup \{i\}} \varrho_{k,i'}^{(\ell)}\bigg]
	\label{eq:LMB_r_fusion}
	\\
	\bar{p}_{k,i}^{(\ell)}(\cdot) & = &
	\sum_{i'\in\mathcal{N}^i \cup \{i\}} \varrho_{k,i'}^{(\ell)}\ p_{k,i'}^{(\ell)}(\cdot)\,\bigg/\, \sum_{i'\in\mathcal{N}^i \cup \{i\}} \varrho_{k,i'}^{(\ell)}
	\label{eq:LMB_p_fusion}
\end{eqnarray}
where 
\begin{equation}
\varrho_{k,i}^{(\ell)}= {r_{k,i}^{(\ell)}}\big/\big[{1-r_{k,i}^{(\ell)}}\big].
\label{eq:varrho}	
\end{equation}

\begin{remark}
	For each object label $\ell$, the fused probability of existence given by~\eqref{eq:LMB_r_fusion} approaches one if one of the $r_{k,i}$ values is close to 1.
	\label{rem:emphasis-on-one}
\end{remark}

The observation made in Remark~\ref{rem:emphasis-on-one} demonstrates the \textit{complementary} nature of information fusion. The safety-critical nature of connected vehicle system technology necessitates comprehensive situational awareness, and avoiding false negatives (miss-detection) is a high priority. This is surmounted by the limited FoV of sensors onboard the vehicles, and the complementary fusion method is hence, a suitable solution.

\subsection{Consensus}
Despite the complementary nature of fusion and its power of inclusion, still the resulting multi-object posterior would be updated up to one layer of neighbors around each vehicle. To ensure a more comprehensive situational awareness is achieved, we propose that during each filtering step (from each time $k$ to $k+1$), the information gathered by vehicle nodes are iteratively propagated through multiple layers of neighbors to each node. 

The most common approach for information propagation in distributed networks is called the \textit{consensus} approach~\cite{4118472}. Consider each node $i$, with a set of neighbors $j\in\mathcal{N}^i$. Let us denote the index to consensus iterations by $\xi$, and in general, consider a parameter estimate $\hat{\theta}_{k,i}$ at node $i$ (after fusion). In the original \textit{consensus} process, the parameter estimate is iteratively updated as follows:
\begin{equation}
    \hat{\theta}_{k,i}(\xi)= \sum_{i'\in\mathcal{N}^i}\omega_{i,i'}\hat{\theta}_{k,i}(\xi-1), 
\end{equation}
where $\omega_{i,i'}$ denotes normalized importance weights for neighboring nodes ($\sum_{i'\in\mathcal{N}^i}\omega_{i,i'} = 1$), not to be confused with particle weights ($w_{k,i,j}^{(\ell)}$). The iterations stop when the change in estimate becomes negligible.

In the proposed solution, we apply the consensus approach with two specific variations. Firstly, instead of a convergence criterion for stopping the iterations, we iterate information propagation for a fixed number of times. This parameter is the practical radius of coverage (number of layers of coverage in the network around each vehicle). Secondly, we do not discriminate between the neighboring vehicles and assign equal weights.

\begin{remark}
	To clarify, let us assume that information exchange occurs for five consensus iterations, and each vehicle has a FoV with 40\,m range around it. This means that after completion of the consensus iterations (each including complementary fusion of the multi-object densities received from neighbors), each vehicle will have a map of environment around it that extends to 40\, m beyond five vehicles around it.
\end{remark}

\section{SMC Implementation}
Consider the vehicle node $i$ in the network and its locally filtered multi-object posterior at time $k$, denoted by $\bm{\pi}_{i,k}$ in which the density of each object state with label $\ell$ is approximated by particles and weights $\left\{\left(w_{k,i,j}^{(\ell)},x_{k,i,j}^{(\ell)}\right)\right\}_{j=1:J_{k,i}^{(\ell)}}.$

Implementation of complementary fusion in each vehicle node is straightforward for probabilities of existence. First the $\varrho_{k,i}^{(\ell)}$ terms are computed as in equation~\eqref{eq:varrho}, then the fused probability of existence is calculated via equation~\eqref{eq:LMB_r_fusion}. Implementing the fusion of single-object densities via~\eqref{eq:LMB_p_fusion} means that at each node $i$, for each object label $\ell$, all the particles at node $i$ and its neighbors $i'\in \mathcal{N}^i$ are concatenated, and their weights are rescaled by a factor of $\varrho_{k,i'}^{(\ell)}$ and normalized. To keep the operation computationally tractable, the large ensemble of concatenated particles are resampled.

The consensus process is implemented as through iterations $\xi = 1:N_{\text{consensus}}$ during the sampling time period between $k$ and $k+1$. In each iteration $\xi$, the vehicle node $i$ receives the local posteriors $\bm{\pi}_{i',k}(\xi)$ from its neighbors and fuses them together with its own local posterior $\bm{\pi}_{i,k}(\xi)$. The result will be $\bm{\pi}_{i,k}(\xi+1)$.

\subsection{Tracking Challenges}

The dynamic and distributed nature of the vehicle network brings in some challenges related to tracking of objects as part of situational awareness. Speaking of \textit{tracking}, we emphasize on the labels of objects and accurate propagation of those labels through the fusion and consensus processes running in each vehicle/sensor node.

\begin{figure}[t]
	\centering
	\includegraphics[width=0.60\linewidth]{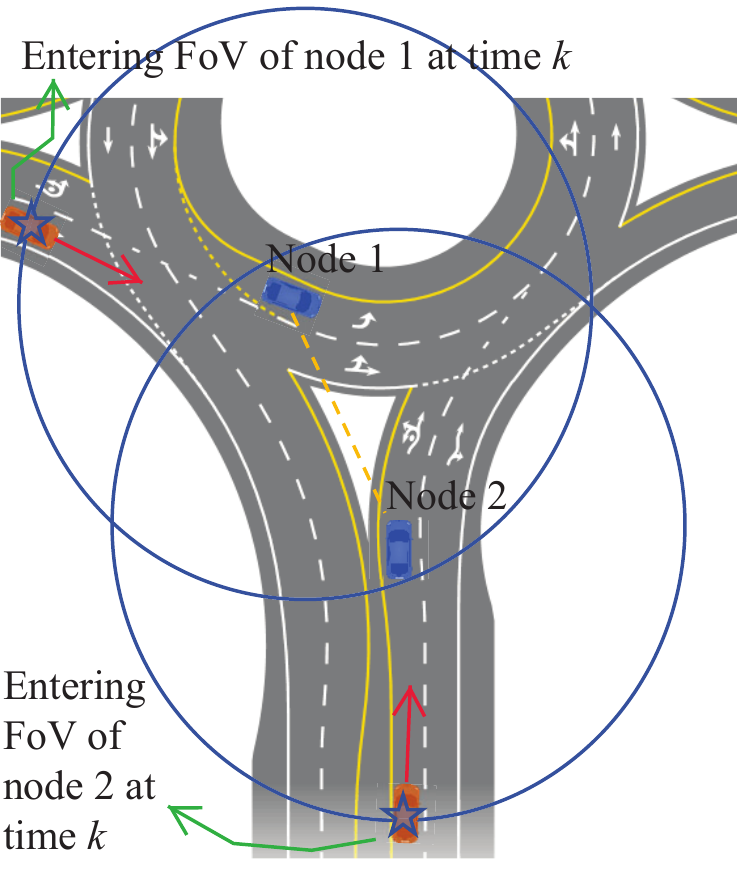}
	\caption{An example of label inconsistency in the proposed distributed fusion scheme. Circles represent each node's FoV. Two sensor nodes in the network (blue) each detect a vehicle (orange) that is entering their FoV at the same time $k$. The orange vehicles are different objects but may be assigned the same label $(k,m)$.}
	\label{fig:label_inconsistency}
\end{figure}

The repeated unification of all Bernoulli components through complementary fusion through consensus iterations, and the movements of vehicles in the network can lead to two major challenges. The first challenge is the possibility of two different objects being assigned the same label. An example is shown in Fig.~\ref{fig:label_inconsistency} where two different objects enter the FoV of two different vehicle nodes at the same time $k$. Hence, as part of the birth process locally running at each node, they may be both labeled as $(k,m)$, and their probabilities of existence will increase due to their detection within the FoV of each node. When the resulting LMB posteriors at nodes 1 and 2 are fused, the label inconsistency will cause drastic errors treating two separate objects as one.

Our proposed solution is straightforward and very intuitive. We propose that at each node, the node identity is appended to the labels of newly detected objects, i.e. object labels are formed as a triple $(k,i,m)$ where $i$ is the identity of the sensor node where the object was detected for the first time.

\begin{figure}[t]
	\centering
	\includegraphics[width=0.67\linewidth]{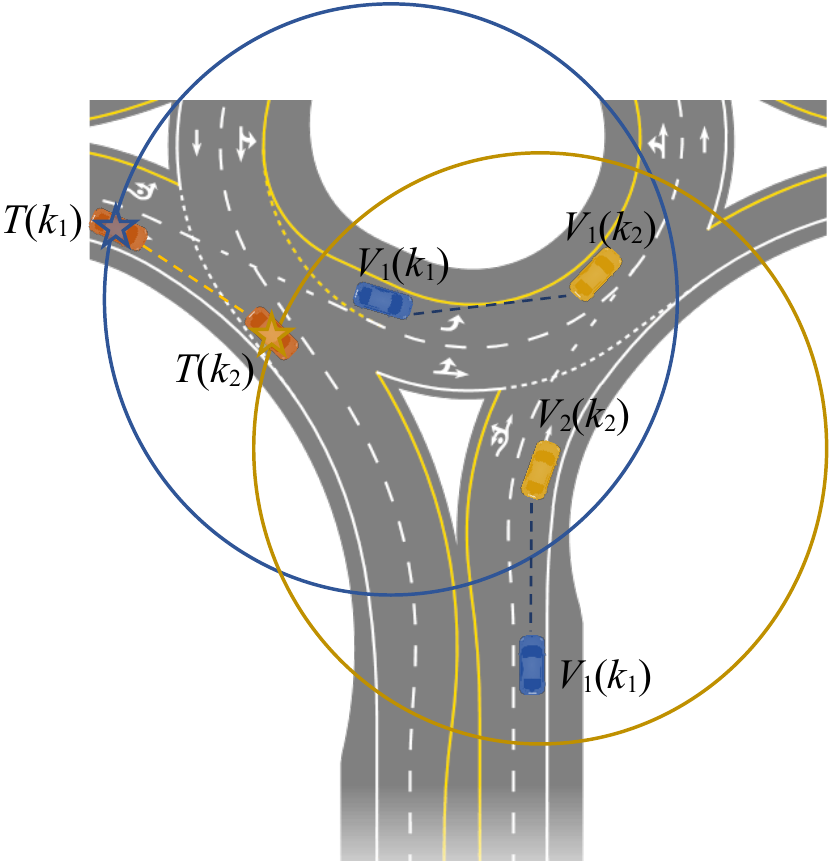} 
	\caption{An example of double counting in a distributed RFS network (top) and label inconsistency (bottom). Circles represent each nodes FoV. The double counting illustration highlights the potential for a nodes posterior to have multiple labels belonging to the same target. The evolution between two times is shown, with an orange target vehicle moving between $k_1$ and $k_2$. Blue vehicles represent vehicles at the first time, while yellow is their evolution at the second. The bottom shows two vehicles (blue) each running independent RFS filters detecting a target vehicle (orange) at the same time and assigning the same label to each, which when communicated presents inconsistency in target information.}
	\label{fig:double_counting}
\end{figure}

Another major challenge is that the complementary fusion scheme is designed to be comprehensive and the occurrence of information \textit{double-counting} is very likely. In particular, we may end up with multiple labels assigned to the same object. An example is shown in Fig.~\ref{fig:double_counting}. 

An object $T$ moves from time $k_1$ till time $k_2$. At time $k_1$, it enters the FoV of vehicle node $V_1$ (with identity $i=1$) and is detected as a newly born object with label $(k_1,1,m_1)$ where $m_1$ can be any number from 1 to the number of birth components in the local filter running at the node.  Later at time $k_2$, the same object enters the FoV of vehicle node $V_2$ ($i=2$) and is given a new label of $(k_2,2,m_2)$. This is while, in the LMB posterior that is formed in node $V_1$ and communicated and fused in node $V_2$, the same object is represented with a Bernoulli component labeled $(k_1,1,m_1)$. 

Our remedy for the double-counting issue is formulated based on the fact that same object represented with two different labels is expected to have density particles that are centered very close to each other. Indeed, by investigating the distance between the EAP state estimates of the two objects, we can merge them into one if the distance is below a small threshold. 

When the two Bernoulli components with different labels are merged, their labels, probabilities of existence, and state densities need to combine. The merged label can be simply chosen as the \textit{older} label. The probabilities of existence can be combined according to the union rule: sum of the two minus their product. The densities are merged by concatenating the particles with their weights being normalized. We propose to scale the weights of particles of each Bernoulli component by its probability of existence, i.e. giving more emphasis on the particles of the Bernoulli component that is more likely to exist. Obviously, the particles need to be resampled at the end, to maintain computational tractability. Algorithm~\ref{alg:merging} presents the step-by-step pseudocode for the merging operation. 

\begin{algorithm}[t]
	\caption{Merging of Bernoulli components.}
	\label{alg:merging}
	\begin{algorithmic}[1]
		\Statex \underline{\textsc{Inputs:}} $\ell_1, \ell_2, r_{k,i}^{(\ell_1)}, r_{k,i}^{(\ell_2)}$, $\left\{\left(w_{k,i,j}^{(\ell_1)},x_{k,i,j}^{(\ell_1)}\right)\right\}_{j=1:J_{k,i}^{(\ell_1)}}$
		,
		$\left\{\left(w_{k,i,j}^{(\ell_2)},x_{k,i,j}^{(\ell_2)}\right)\right\}_{j=1:J_{k,i}^{(\ell_2)}}$, $\bar{J}$
		
		\Statex
		\Statex\Comment $\bar{J}$: the desired number of particles after merging.
		\Statex\Comment $\ell_1 = (k_1,i_1,m_1),$  $\ell_2 = (k_2,i_2,m_2),$
		$k \geqslant k_1,$ $k \geqslant k_2$
		
		\Statex\Statex \underline{\textsc{Outputs:}} $\bar{\ell}, r_{k,i}^{(\bar{\ell})}$, $\left\{\left(w_{k,i,j}^{(\bar{\ell})},x_{k,i,j}^{(\bar{\ell})}\right)\right\}_{j=1:\bar{J}}$
		
		\Statex \Comment Merged prob. of existence and particles and label.
		\Statex
		
		\If{$k_1<k_2$} 
			 $\bar{\ell} \gets \ell_1$
		\Else 
			 $\ \bar{\ell} \gets \ell_2$
		\EndIf \Comment Merged label is the older one.
		\Statex
		\State $r_{k,i}^{(\bar{\ell})} \gets r_{k,i}^{(\ell_1)} + r_{k,i}^{(\ell_2)} - r_{k,i}^{(\ell_1)}r_{k,i}^{(\ell_2)}$
		\Statex \Comment Merged prob. of existence computed by union rule.
		\State $\alpha_1 \gets {r_{k,i}^{(\ell_1)}}\bigg/ \left(r_{k,i}^{(\ell_1)}+r_{k,i}^{(\ell_2)}\right);$
		\ $\alpha_2 \gets \left(1-\alpha_1\right)$ 
		\Statex
		\Comment Scale factors set for particles of $(\ell_1)$ and $\ell_2$
		\State $\bm{x} \gets \emptyset$ 
		\Comment $\bm{x}$ to contain all pairs of particles and weights.
		\For{$j=1:J_{k,i}^{(\ell_1)}$}
			 \ \ \ \ $\bm{x} \gets \bm{x}\cup \left\{\left(\alpha_1\,w_{k,i,j}^{(\ell_1)},x_{k,i,j}^{(\ell_1)}\right)\right\}$
		\EndFor
		\For{$j=1:J_{k,i}^{(\ell_2)}$}
			\ \ \ \ $\bm{x} \gets \bm{x}\cup \left\{\left(\alpha_2\,w_{k,i,j}^{(\ell_2)},x_{k,i,j}^{(\ell_2)}\right)\right\}$
		\EndFor
		\Statex
		\State $\left\{x_{k,i,j}^{(\bar{\ell})}\right\}_{j=1:\bar{J}} =$
		 \Call{Resample}{$\bm{x},\bar{J}$}
		\For{$j = 1: \bar{J}$}\ \ $w_{k,i,j}^{(\bar{\ell})} \gets {1}\big/{\bar{J}}$
		\EndFor
		\Statex \Comment Concatenated particles are resampled into $\bar{J}$ equally 
		\Statex \hspace{7.5mm}weighted particles.
	\end{algorithmic}
\end{algorithm}

\section{Simulation Results}
\begin{figure}[t]
	\centering
	\includegraphics[width=0.80\linewidth]{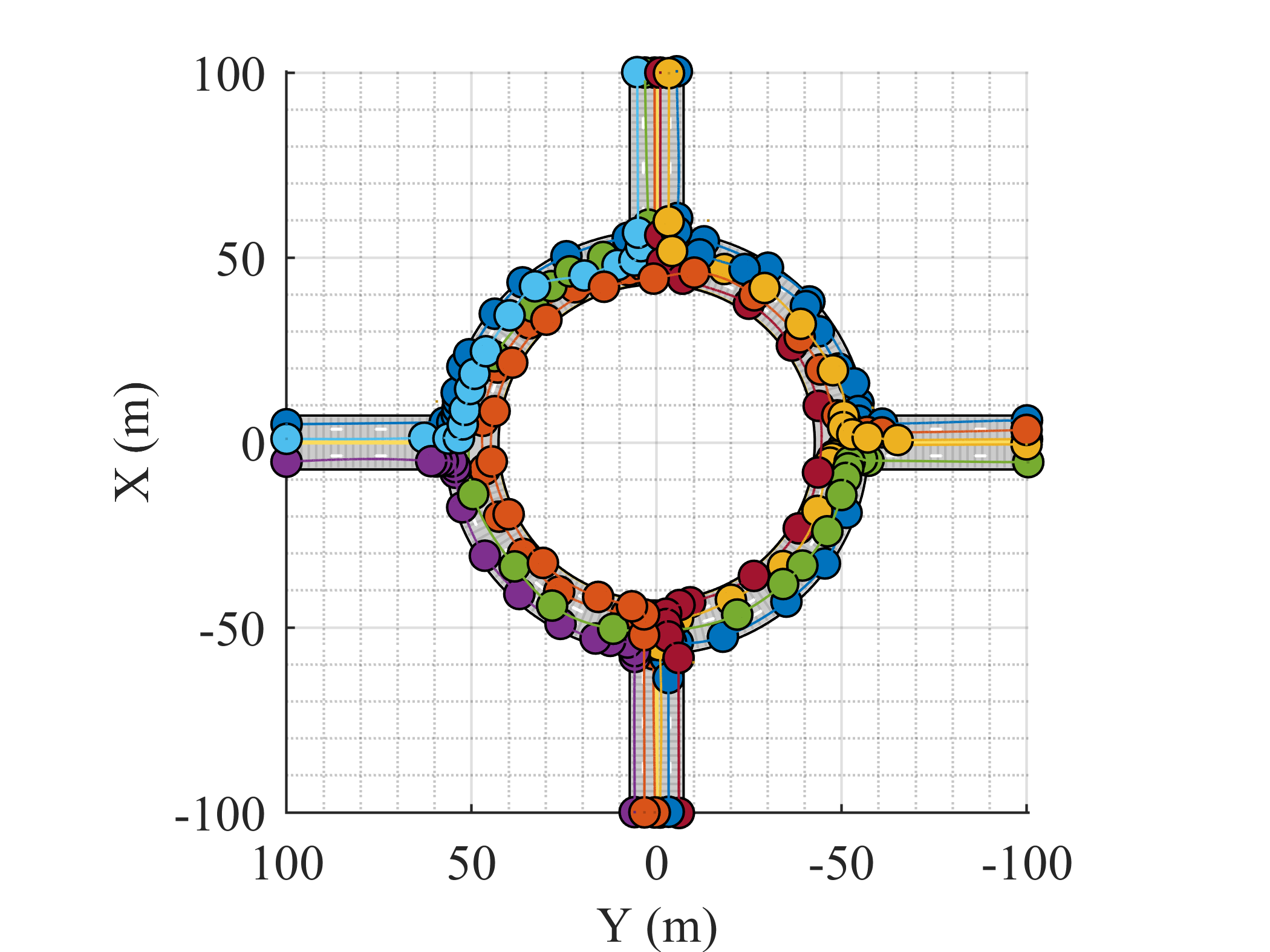}
	\caption{Setup of the simulation. Colored lines represent individual trajectories, which are defined by their respective coloured waypoints.}
	\label{fig:scenario}
\end{figure}

We evaluated the performance of our proposed distributed fusion method using the driving scenario designer application in MATLAB. The scenario contains 10 vehicles following a realistic and complex trajectory including sharp turns with large acceleration variations. The scenario encompasses a 200\,m\,$\times$\,200\,m area, as shown in Fig.~\ref{fig:scenario}. 

Each vehicle has a single radar sensor with a limited field of view. Mobile radars have 50\,m range. There is also a stationary radar with a range of 120\,m. All sensors have 360 degree coverage. Vehicles are modeled as 0.1\,cm$\times$0.1\,cm points to ensure that a single measurement is generated per sensor. Each vehicle's state includes its position, velocity and turning rate: $x = [p_x\ \,p_y\ \,\dot{p}_x\ \,\dot{p}_y\ \,\omega]^\top$. Motion was approximated by a constant turn (CT) model with a 0.1 second sampling period. In CT model, the single-object state transition density is given by $
f_{k|k-1}(\cdot|x) = \mathcal{N}(\cdot; m(x),Q)
$ 
where 
$
m(x)=[(F(\omega)\times x)^\top \,\ \omega]^\top,
$ 
and
${Q}=\text{diag}\big(\sigma^2_\omega G G^\top, \sigma^2_u\big)$
in which $\sigma_\omega=15$\,m/s\textsuperscript{2} and $\sigma_u=30^\circ$/s, and
$$
    F(\omega)=\begin{bmatrix}
    1&\frac{\sin{\omega}}{\omega}&0&-\frac{1-\cos{\omega}}{\omega}\\
    0&\cos{\omega}&0&-\sin{\omega}\\
    0&\frac{1-\cos{\omega}}{\omega}&1&\frac{\sin{\omega}}{\omega}\\
    0&\sin{\omega}&0&\cos{\omega}\end{bmatrix},
    G=\begin{bmatrix}
    \frac{1}{2}&0\\
    1&0\\
    0& \frac{1}{2}\\
    0&1
    \end{bmatrix}.
$$
Finally, for each sensor, object birth is modelled as an LMB with its elements covering the rim of the detection range.

The dynamic nature of scenario entails the vehicles entering and exiting the scene at various times. The ground-truth times are listed in  Table~\ref{table:birth_death}. 

\begin{table}[t]
\caption{Ground-truth entry (birth) and exit (death) times}
\label{table:birth_death}
\begin{center}
	\begin{tabular}{rrrrrr}
		\hline
		\textbf{Veh. no.} & $\bm{k}_{\text{\textbf{birth}}}$ & $\bm{k}_{\text{\textbf{death}}}$ & \textbf{Veh. no.} & $\bm{k}_{\text{\textbf{birth}}}$ & $\bm{k}_{\text{\textbf{death}}}$\\
		\hline
		1 & 1 & 232 & 6 & 12 & 153\\
		2 & 32 & 274 & 7 & 12 & 211 \\
		3 & 2 & 224 & 8 & 2 & 193 \\
		4 & 22 & 155 & 9 & 72 & 311 \\
		5 & 52 & 289 & 10 & 52 & 205 \\
		\hline
	\end{tabular}
\end{center}
\end{table}

\begin{figure}[ht]
    \centering
    \includegraphics[width=0.80\linewidth]{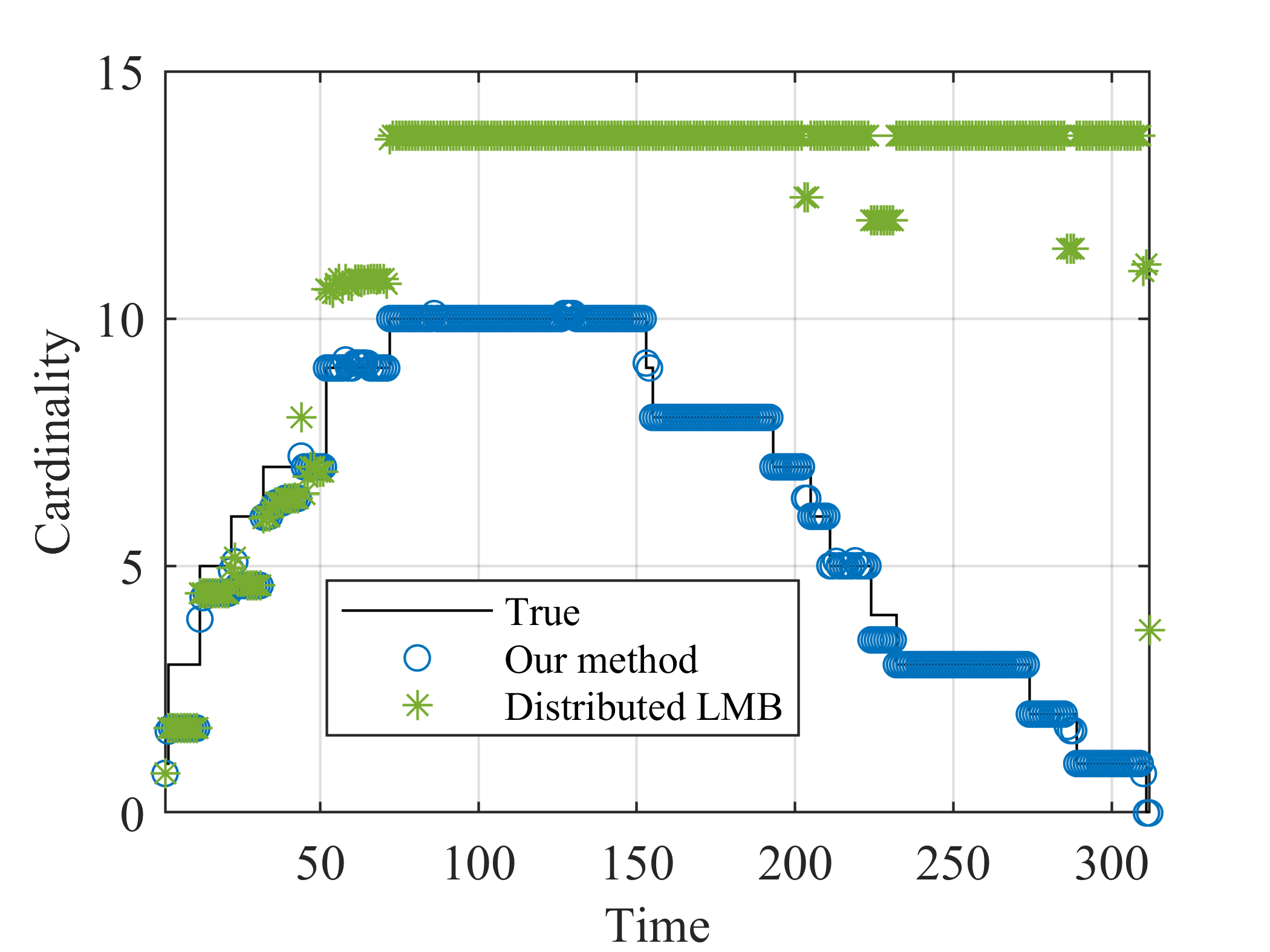}
    \caption{Cardinality estimates for 10 iterations averaged across all sensors. 3 Consensus Iterations.}
    \label{fig:cardinality}
\end{figure}

\begin{figure}[ht]
    \centering
  
    \includegraphics[width=0.80\linewidth]{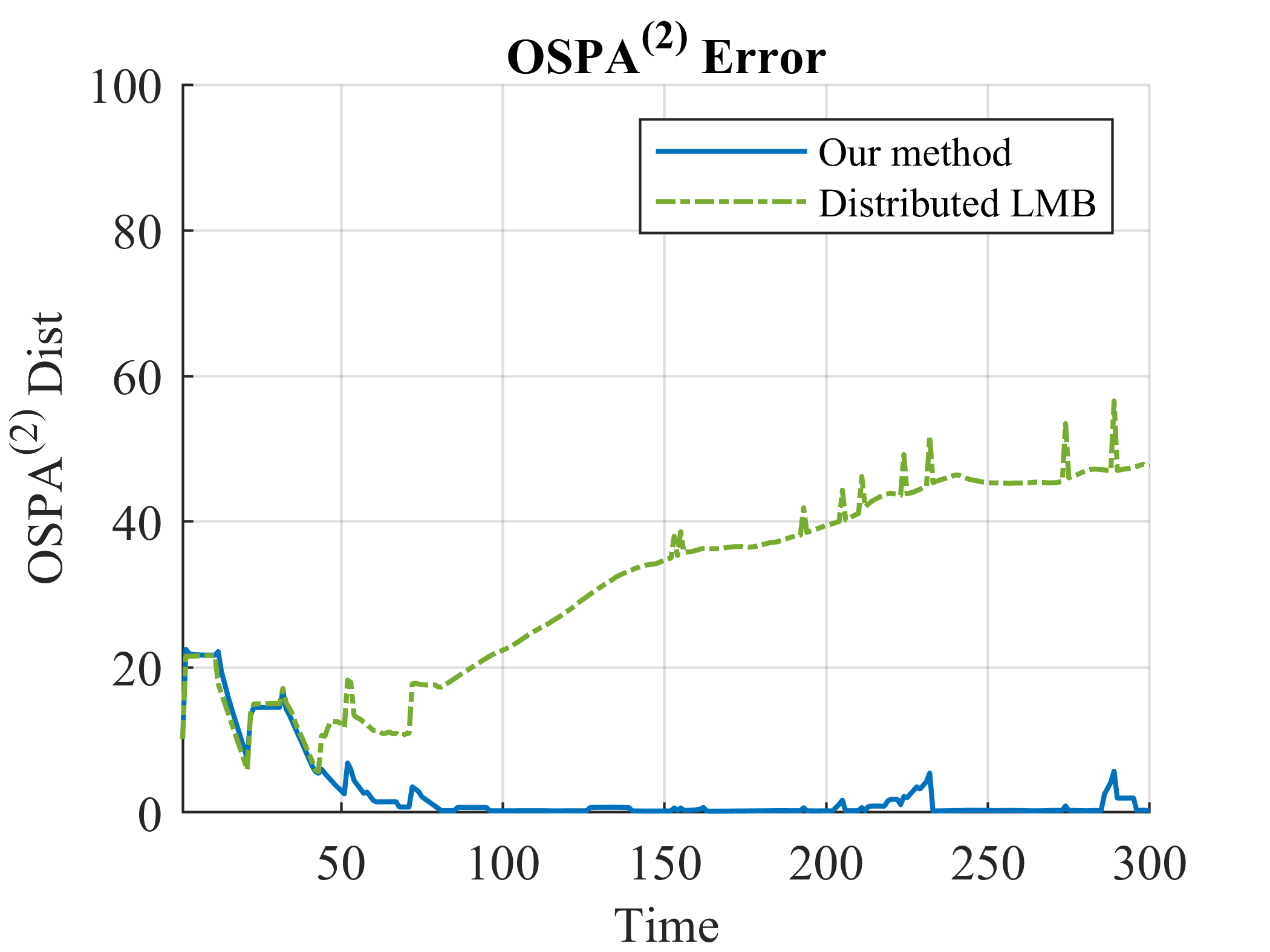}
    \caption{OSPA$^{(2)}$ estimates for 10 iterations averaged across all sensors for a window length of 10, with order 1 and cutoff 50. 3 Consensus Iterations.}
    \label{fig:ospa2}
\end{figure}

\begin{figure}[ht]
    \centering
   \includegraphics[width=0.80\linewidth]{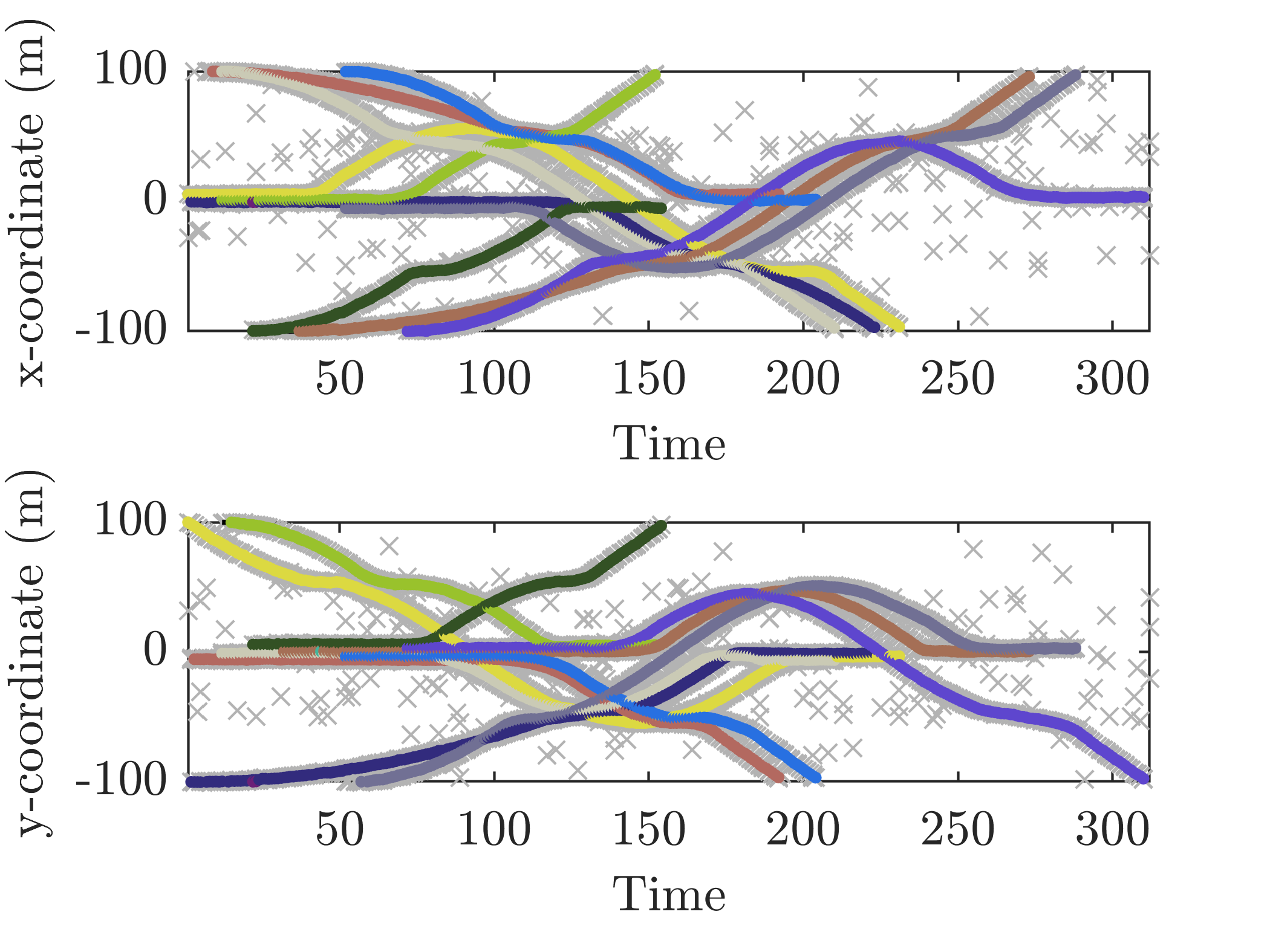}
    \caption{track estimates estimates for a single iteration. 3 Consensus Iterations.}
    \label{fig:tracks}
\end{figure}

In our experiments, we examined the distributed complementary fusion with consensus once without and once with our proposed solutions for tracking challenges. In the former case, we use the term ``Distributed LMB'', and in the latter, ``Our method'' in the figure legends.

Figure~\ref{fig:cardinality} shows the cardinality (number of objects returned in estimates extracted from local LMB posterior) values averaged across all sensors in the network. Evidently, due to double-counting and label inconsistency, objects are included with multiple labels and with original label, there will be different objects with the same labels. The result (green * points in Fig.~\ref{fig:cardinality}) shows a persisting overestimate of the number of objects. This is while after putting our proposed extension of labels (to include sensor identities) and merging algorithm, almost all the time, we get accurate cardinality estimates.

The impact of cardinality overestimation on OSPA$^{(2)}$ errors~\cite{8217598} is demonstrated in Fig.~\ref{fig:ospa2}. The error is significantly reduced when our method is applied. The tracking results of the proposed method are presented for a single iteration in Fig.~\ref{fig:tracks}. We can see that despite the reasonable amount of clutter (representing false alarms and other objects in the scene) for an ITS scenario, each object is successfully identified and tracked for the entire duration of the simulation. Tracks are not lost, although there are occasionally false tracks generated which are quickly killed off. Despite the initial spike in OSPA$^{(2)}$ (Fig.~\ref{fig:ospa2}), showing a delay in obtaining the tracks, measured over a window of 10 steps (1 second), once they are obtained, they show stable performance, except at the spike between $k=200$ and $k=250$, as well as around $k=300$. This is due to targets leaving the scene and the subsequent cardinality error.

\section{Conclusion}
This paper presented a new distributed information fusion algorithm using a consensus-based complementary fusion method. A novel solution to solving the double-counting and label inconsistency problems was introduced and  shown to perform with satisfactory in a complex connected driving scenario.

\section*{Acknowledgment}
This work was supported by the Australian Research Council through Grant DE210101181.

\bibliographystyle{ieeetr}
\bibliography{bibliography}
\vspace{12pt}
\color{red}

\end{document}